\documentclass[10pt,aps,prd,noshowpacs,eqsecnum,nofootinbib]{revtex4-1}
\usepackage{amsmath,amsthm,amssymb}
\usepackage{mathrsfs,bbm}

\usepackage{nicefrac} 

\usepackage{longtable}

\usepackage{latexsym}

\usepackage{array}

\usepackage{color}

\begin{document}


\title{Texture Zeros and WB Transformations in the Quark Sector of the Standard Model }



\author{Yithsbey Giraldo}
\affiliation{Departamento de F\'\i sica, Universidad de Nari\~no, A.A. 1175, San Juan de Pasto, Colombia}
\date{\today}

\begin{abstract}
Stimulated by the recent attention given to the  texture zeros found in the quark mass matrices sector of the Standard Model,
an analytical method for identifying (or to exclude)  texture zeros models will be implemented here,
{starting from arbitrary quark mass matrices and making a suitable weak
basis~(WB) transformation}, we are be able to find equivalent quark mass matrix. It is shown that the number of non-equivalent quark mass matrix representations is finite. We give {exact} numerical results for parallel and non-parallel four-texture zeros  models. We find that some five-texture zeros Ans$\ddot{\textrm{a}}$tze are in agreement with all present experimental data. And we confirm definitely that six-texture zeros of Hermitian quark mass matrices  are not viable models anymore.
\end{abstract}


\maketitle


\section{Introduction}
Although the gauge sector of the Standard Model (SM) with the $SU(3)_C\otimes SU(2)_L\otimes U(1)_Y$  symmetry is very successful,  the Yukawa sector of the SM is still poorly understood. The origin of the fermion masses, the mixing angles and the CP violation remain as open problems in particle physics. There have been a lot of studies of possible fundamental symmetries in the Yukawa  coupling matrices of the 
  SM~\cite{c_1,c_3,c_4}. In the absence of a more fundamental theory of interactions, an independent phenomenological model approach to search for possible textures or symmetries in the fermion mass matrices  is still playing an important role.

In the SM, the mass term is given by
\begin{equation}
 -{\cal L}_M=\bar u_RM_u u_L+\bar d_RM_dd_L+h.c,
\end{equation}
where the mass matrices  $M_u$ and $M_d$ are three-dimensional complex matrices. In the most general case, they contain  36 real parameters. A first simplification, without losing generality, is  by making use of the polar decomposition theorem of matrix algebra, by which, one can always express a general mass matrix as a product of a hermitian and unitary matrix. Therefore, we can consider quark mass matrices to be hermitian as the unitary matrix can be absorbed in the right handed quark fields. This immediately brings down the number of free parameters from 36 to 18.

A simple and instructive ansatz of hermitian quark mass matrices with six-texture zeros   was first proposed in reference~\cite{c_1}.
An additional non-parallel six-texture zeros  was given in~\cite{b_2}. Both textures are currently ruled out~\cite{b4}, 	
because, among other things, they do not reproduce some entries of the Cabibbo-Kobayashi-Maskawa (CKM)  mixing matrix  $V$. Specifically, in both cases,
the magnitude of $|V_{ub}/V_{cb}|$ predicted by $\sqrt{m_u/m_c}$ is too low $(V_{ub}/V_{cb}\approx0.06$ or smaller for reasonable values of the quark masses $m_u$ and $m_c$~\cite{b3,b1}) to agree with the present experimental result ($|V_{ub}/V_{cb}|_{\text{ex}}\approx0.09$~\cite{b3}).
Because of this, some authors have highly recommended the use of  four-texture zeros~\cite{b4,b0,b_1}. {It is shown in this work, that four texture zeros is readily feasible, and, we can even get five-zeros textures.}


We would therefore present an analytical method to calculate models containing various  texture zeros in the quark mass matrix sector, taking into account the latest experimental data provided~\cite{b3}. We use simultaneously, in our research, two very common approach: one approach consists  of placing zeros (called  texture zeros) at certain entries of quark mass matrices that can predict self-consistent and experimentally-favored relations between quark masses and flavor mixing parameters~\cite{b_2,c_5,c_8}; which is used in conjunction with the other approach, the WB transformation (weak basis transformation), that transforms the quark mass matrix representations into new equivalent ones~\cite{b0}.

This paper is organized as follows: in Sect.~\ref{sII} we discuss some issues related to the WB transformation method and its utilities. We dedicate, in Sect.~\ref{s3}, to obtain some numerical parallel and non-parallel four-texture zeros quark mass matrices using special techniques for that; which we use, in Sect.~\ref{sIV}, to find five-texture zeros in quark mass matrices compatible with the present experimental data; this configuration, is studied from an analytical point of view, in Sect.~\ref{sV}; and our conclusions are presented in Sect.~\ref{sVI}.  And the method used extensively throughout this paper to find texture zeros   is verified in Appendix~\ref{a2}.


\section{WB Transformations}
\label{sII}
The most general WB transformation~\cite{b0}, that leaves the physical content invariable and the mass matrices Hermitian, is
\begin{equation}
\label{2aa}
 \begin{split}
  M_u&\longrightarrow M_u^\prime=U^\dag M_u U,\\
M_d&\longrightarrow M_d^\prime=U^\dag M_d U,
 \end{split}
\end{equation}
where $U$ is an arbitrary unitary matrix. We say that the two representations $M_{u,d}$ and $M_{u,d}^\prime$ are equivalent each other.  Besides, it implies that the number of equivalent representations is  infinity. 
This kind of transformation  will be used extensively in calculations below. 

But, firstly, let us show  that the  WB  transformation is  exhaustive in generating all possible mass matrix representations.  Let us first consider the representation of Hermitian quark mass matrices indicated by $(M_u,M_d)$, and diagonalize them as follows
\begin{equation}
\label{19}
 U_u^\dag M_u U_u=D_u\quad\textrm{and}\quad U_d^\dag M_dU_d=D_d.
\end{equation}
The CKM mixing matrix is given by
\begin{equation}
\label{20}
 V_{ckm}=U_u^\dag U_d.
\end{equation}
On the other hand, the prime representation $(M^\prime_u,M^\prime_d)$ gives
\begin{equation}
\label{21}
 U_u^{\prime\dag} M^\prime_u U^\prime_u=D_u\quad\textrm{and}\quad U_d^{\prime\dag} M^\prime_dU^\prime_d=D_d,
\end{equation}
and
\begin{equation}
\label{22}
 V_{ckm}=U_u^{\prime\dag} U^\prime_d.
\end{equation}
Equating the expressions~(\ref{20}) and~(\ref{22}) yields
\begin{equation}
\label{23}
 U_u^\dag U_d=U_u^{\prime\dag} U^\prime_d\Rightarrow U^\prime_uU_u^\dag=U_d^\prime U_d^\dag.
\end{equation}
And equating~(\ref{19}) and~(\ref{21}), gives respectively
\begin{equation}
 U_u^{\prime\dag}M_u^\prime U_u^\prime=U_u^\dag M_u U_u\quad\textrm{and}\quad U_d^{\prime\dag}M_d^\prime U_d^\prime=U_d^\dag M_d U_d,
\end{equation}
where we find that the mass matrices $M_u$ and $M_d$ can be expressed in terms of the mass matrices $M_u^\prime$ and $M_d^\prime$ as follows
\begin{align}
\label{A7}
 M_u&=U_uU_u^{\prime\dag}M_u^\prime U_u^\prime U_u^\dag,\\
\label{26}
M_d&=U_dU_d^{\prime\dag}M_d^\prime U_d^\prime U_d^\dag.
\end{align}
Using~(\ref{23}) into~(\ref{26}), we have
\begin{equation}
\label{A9}
 M_d=U_uU_u^{\prime\dag}M_d^\prime U_u^\prime U_u^\dag.
\end{equation}
where $U=U_uU_u^{\prime\dag}$ is an unitary matrix which allows us to state.

\medskip
\begin{equation}
\label{2.11a}
\begin{split}
&\textit{ \small ``In the SM, any two pairs of Hermitian quark mass matrices, given by $(M_u,M_d)$ and $(M^\prime_u,M^\prime_d)$, with identical} \\
&\textit{\small  eigenvalues and {flavor mixing parameters}, to a specific scale energy, are related through a WB transformation,''}
\end{split}
\end{equation}
%
i.e., there is no  a quark mass  matrix representation outside the set~(\ref{2aa}). In this reasoning, we have assumed that both representations generates {the same entries, including the phases}, for the CKM mixing matrix~($ V_{ckm}$); something valid due that a WB transformation makes them equal, as will be shown in section~(\ref{3_2}).
%

%

The importance of the WB transformation, as  calculation tool, can be appreciated from the following results.

%

\subsection{The preliminary matrix representation}
\label{3_2}
In the quark-family basis, it is more convenient to use the following quark mass matrix representation~\cite{b0,b2}
\begin{equation}
\label{3}
\begin{split}
 M_u&=D_u=\begin{pmatrix}
          \lambda_{1u}&0&0\\
0&\lambda_{2u}&0\\
0&0&\lambda_{3u}
         \end{pmatrix},\\
%
 M_d&=VD_dV^\dag,
\end{split}
\end{equation}
which comes from a WB transformation, and we call it as the \textit{the u-diagonal representation}. We call the other possibility
\begin{equation}
\label{4}
\begin{split}
 M_u&=V^\dag D_uV,\\
%
 M_d&=D_d=\begin{pmatrix}
          \lambda_{1d}&0&0\\
0&\lambda_{2d}&0\\
0&0&\lambda_{3d}
         \end{pmatrix},
\end{split}
\end{equation}
as  \textit{the d-diagonal representation}. One advantage of using representations~(\ref{3})~(or~(\ref{4})) is to be able to use simultaneously the CKM mixing matrix $V$ and the quark mass eigenvalues  $|\lambda_{iu,d}|$ ($i=1,2,3$). Where $\lambda_{iu,d}$ may be either positive or negative and satisfy the hierarchy
\begin{equation}
\label{19d}
|\lambda_{1u,d}|\ll|\lambda_{2u,d}|\ll|\lambda_{3u,d}|.
\end{equation}
%
%



%
It is usually said that the CKM matrix is an arbitrary unitary matrix with five phases rotated away through the phase redefinition of the left handed up and down quark fields~\cite{b2p}. This can be shown by using the following unitary matrix 
\[
\begin{pmatrix}
    e^{ix}&&\\
&e^{iy}&\\
&&1
\end{pmatrix}
\]
in order to make a WB transformation on~(\ref{3}). The up matrix
\begin{equation}
\label{3_10}
M_u=\begin{pmatrix}
 e^{ix}&&\\
&e^{iy}&\\
&&1
\end{pmatrix} D_u \begin{pmatrix}
 e^{ix}&&\\
&e^{iy}&\\
&&1
\end{pmatrix}^\dag=D_u,
\end{equation}
remains equal, while the down matrix takes the form
\begin{align}
\label{21b}
 M_d&=\begin{pmatrix}
 e^{ix}&&\\
&e^{iy}&\\
&&1
\end{pmatrix}\left(V 
D_dV^\dag\right)\begin{pmatrix}
 e^{ix}&&\\
&e^{iy}&\\
&&1
\end{pmatrix}^\dag,\\
\label{22b}
\begin{split}
M_d&=\left[\begin{pmatrix}
 e^{ix}&&\\
&e^{iy}&\\
&&1
\end{pmatrix}V\begin{pmatrix}
 e^{i\alpha_{1}}&&\\
&e^{i\alpha_{2}}&\\
&&e^{i\alpha_{3}}
\end{pmatrix}\right] 
D_d 
\left[\begin{pmatrix}
 e^{ix}&&\\
&e^{iy}&\\
&&1
\end{pmatrix}V\begin{pmatrix}
 e^{i\alpha_{1}}&&\\
&e^{i\alpha_{2}}&\\
&&e^{i\alpha_{3}}
\end{pmatrix}\right]^\dag,
\end{split}
\end{align}
where in the last step we have used the identity~(\ref{3_10}) applied to the diagonal down mass  matrix.
The expression into the square brackets is precisely the most general way to write an unitary matrix~\cite{b2p}.

In this representation, the  matrix $M_d$, in~(\ref{21b}), contains two free parameters $x$  and $y$, which plays an important role to obtain   texture  zeros as we shall see later.

\subsection{A unique negative eigenvalue}

The result~(\ref{2.11a}) permits us to use the u-diagonal representation~(\ref{3}) (or the d-diagonal representation~(\ref{4})) as the starting point, to generate any other representation. If they exist, by this method, important texture zeros in mass matrix can be found.

Because some texture zeros must lie  along its diagonal entries of both up and down Hermitian  quark mass matrices, it implies that at least one and at most two of its eigenvalues must be negative~\cite{b0}.
Furthermore, for the case of two negative eigenvalues, these mass matrices can be reduced to have only  one negative eigenvalue, by factoring a minus sign out which can be included, for instance, into the mass matrix basis~(\ref{3}). Thus, without loss of generality, the texture zeros models can be deduced considering that

\medskip
\begin{equation}
\label{2.9}
\begin{split}
&\textit{ ``each one of quark mass matrices} \; M_u \; and\;  M_d\\ 
&\textit{contains exactly one negative eigenvalue.''}
\end{split}
\end{equation}


\section{Numerical Four-Texture Zeros }
\label{s3}
There are a wide variety of four-texture zeros representations. Using a specific approach, some non-parallel texture are easy to obtain. But more laborious methods  are required in  parallel cases. In our analysis we will use the next physical quantities.


\subsection{Quark masses and CKM}
\label{A}
{For quark mass matrix phenomenology, values of $m_q (\mu)$ at $\mu = m_Z$ are useful, because
the observed CKM matrix parameters $|V_{ij} |$ are given at $\mu = m_Z$ . We summarize quark masses at $\mu = m_Z$}~\cite{b3a,b1,b2}.

%
{
{\small
\begin{equation}
\label{17a}
\begin{split}
 m_u&=1.38^{+0.42}_{-0.41}\,,\: m_c=638^{+43}_{-84},\: m_t=172100\pm{1200}\,,\\
m_d&=2.82\pm0.48\,,\: m_s=57^{+18}_{-12}\,,\: m_b=2860^{+160}_{-60}.
\end{split}
\end{equation}}}
given in units of MeV.
%

The Cabibbo-Kobayashi-Maskawa~(CKM) matrix~\cite{b16,b17,b1} is a $3\times3$ unitary matrix,
\begin{equation*}
 V=\begin{pmatrix}
    V_{ud}&V_{us}&V_{ub}
\\
V_{cd}&V_{cs}&V_{cb}
\\
V_{td}&V_{ts}&V_{tb}
   \end{pmatrix},
\end{equation*}
 which can be parametrized by three mixing angles and the CP-violating Kobayashi-Maskawa~(KM) phase~\cite{b17}. Of the many possible conventions, a standard choice has become~\cite{b18}
\begin{widetext}
{
\begin{equation}
\label{3.2}
 V=\begin{pmatrix}
    c_{12}\,c_{13}&s_{12}\,c_{13}&s_{13}\,e^{-i\delta}\\
-s_{12}\,c_{23}-c_{12}\,s_{23}\,s_{13}\,e^{i\delta}& c_{12}\,c_{23}-s_{12}\,s_{23}\,s_{13}\,e^{i\delta}&
s_{23}\,c_{13}\\
s_{12}\,s_{23}-c_{12}\,c_{23}\,s_{13}\,e^{i\delta}&-c_{12}\,s_{23}-s_{12}\,c_{23}\,s_{13}\,e^{i\delta}&c_{23}\,c_{13}
   \end{pmatrix},
\end{equation}}
\end{widetext}
%
where $s_{ij}=\sin\theta_{ij},c_{ij}=\cos\theta_{ij}$, and $\delta$ is the phase responsible for all CP-violating phenomena in flavor-changing processes in the SM. The angles $\theta_{ij}$ can be chosen to lie in the first quadrant, so $s_{ij},c_{ij}\ge0$.

It is known experimentally that $s_{13}\ll s_{23}\ll s_{12}\ll1$, and it is convenient to exhibit this hierarchy using the Wolfenstein parametrization. We define~\cite{b19,b20} 
\begin{equation}
\label{3.3}
 \begin{split}
  s_{12}&=\lambda,\quad\quad \quad\quad s_{23}=A\,\lambda^2,\\
s_{13}\,e^{i\delta}&=\frac{A\,\lambda^3(\bar\rho+i\,\bar\eta)\sqrt{1-A^2\lambda^4}}{\sqrt{1-\lambda^2}\left[1-A^2\,\lambda^4(\bar\rho+i\,\bar\eta)\right]}.
 \end{split}
\end{equation}
The constraints implied by the unitarity of the three generation CKM matrix
significantly reduce the allowed range of some of the CKM elements. The fit for the
Wolfenstein parameters defined in Eq.~(\ref{3.3}) gives
{
\begin{equation}
\label{3.4}
 \begin{split}
  \lambda&=0.22535\pm0.00065,\quad A=0.811^{+0.022}_{-0.012},\\
\bar\rho&=0.131^{+0.026}_{-0.013},\quad\bar\eta=0.345^{+0.013}_{-0.014}.
 \end{split}
\end{equation}}
These values are obtained using the method of Refs.~\cite{b19,b21}. The fit results for the values of all nine CKM elements are.
{
{\scriptsize
\begin{equation}
\label{2}
V=\begin{pmatrix}
 0.974272 & 0.225349& 0.00351322\,e^{-i\,1.20849} \\
 0.225209\,e^{-i\,3.14101} & 0.97344\,e^{-i\,3.13212\times10^{-5}} & 0.0411845 \\
0.00867944\,e^{-i\,0.377339} & 0.0404125\,e^{-i\,3.12329} & 0.999145
\end{pmatrix},
\end{equation}}
with magnitudes
{\scriptsize
\begin{equation}
\label{2a}
|V|=\begin{pmatrix}
 0.97427\pm0.00015  & 0.22534\pm0.00065& 0.00351^{+0.00015}_{-0.00014} \\
 0.22520\pm0.00065 & 0.97344\pm{0.00016}& 0.0412^{+0.0011}_{-0.0005} \\
0.00867^{+0.00029}_{-0.00031} & 0.0404^{+0.0011}_{-0.0005}& 0.999146^{+0.000021}_{-0.000046}
\end{pmatrix},
\end{equation}}}
%
and the Jarlskog invariant is
\begin{equation}
\label{3.7}
 J=\left(2.96^{+0.20}_{-0.16}\right)\times10^{-5}.
\end{equation}


\subsection{Non-parallel four-texture zeros }
\label{s3.1}
It is the most simple case. For instance, let us take the eigenvalues signs pattern as follow
\begin{align}
 \lambda_{1u}&=-m_u,\lambda_{2u}=m_c,\lambda_{3u}=m_t,\\
\lambda_{1d}&=m_d,\lambda_{2d}=-m_s,\lambda_{3d}=m_b.
\end{align}
Then, for this case, the numerical values in the u-diagonal representation~(\ref{3}) are
%
{
{\scriptsize 
\begin{equation}
\label{20a}
\begin{split}
 M_u&=\begin{pmatrix}
      -1.38&&\\
&638&\\
&&172100
     \end{pmatrix}\textrm{MeV},\\
 M_d&={\begin{pmatrix}
      -0.2\pm0.8 & -12.9758 - 0.386978 i& 4.09941 - 9.38819 i\\
-12.9758 + 0.386978 i & -49.0183 & 119.924- 0.043146i\\
4.09941 + 9.38819i& 119.924 + 0.043146 i& 2855.02
     \end{pmatrix}\textrm{MeV}\,,}
\\
 &={\begin{pmatrix}
      0 & -12.9758 - 0.386978 i& 4.09941 - 9.38819 i\\
-12.9758 + 0.386978 i & -49.0183 & 119.924- 0.043146i\\
4.09941 + 9.38819i& 119.924 + 0.043146 i& 2855.02
     \end{pmatrix}\textrm{MeV}\,,}
\end{split}
\end{equation}}}
%
\hspace{-1mm}where we have used the numerical CKM matrix~(\ref{2}) {and errors of~(\ref{2a}). In the second mass matrix above, in the entry $M_d(1,1)=-0.2\pm0.8$ calculated, since the uncertainty ($\pm0.8$) in determining this element exceeds the value of  0.2 it is obviously reasonable to call the $(1,1)$ entry zero~($M_d(1,1)= 0$).} Something pointed out in Reference~\cite{b2}.


Making a WB transformation on~(\ref{20a}) using the following unitary matrix
\begin{equation}
\label{9}
 U=\begin{pmatrix}
    \cos\theta & 0& \sin\theta\\
0&1&0\\
-\sin\theta &0&\cos\theta
   \end{pmatrix},
\end{equation}
with
%
 $\tan\theta=\sqrt{\frac{m_u}{m_t}}$,
%
the matrices~(\ref{20a}) transform into a form, where the entries  $(1,1)$, $(1,2)$ and $(2,3)$ of matrix $M_u$ becomes zero. Then, we have
%
\begin{equation}
\label{11}
 M_u^\prime=UM_uU^\dag=
\begin{pmatrix}
 0&0&487.338\\
0&638&0\\
487.338 & 0&172099
\end{pmatrix}\textrm{MeV},
\end{equation}
and
%
%
{\scriptsize
\begin{equation}
\begin{split}
 &M_d^\prime=UM_dU^\dag
\\
&=\begin{pmatrix}
 0& -12.6361 - 0.386854i & 12.1844 - 9.38819 i\\
-12.6361 + 0.386854 i &-49.0183 & 119.96 - 0.0442417 i\\
12.1844 + 9.38819i &119.96 + 0.0442417 i & 2854.97
\end{pmatrix}\textrm{MeV},
\end{split}
\end{equation}}
{where the element $M_d^\prime(1,1)$ is zero for the same reason given in~(\ref{20a}).}
We finally obtain a non-parallel four-texture  zeros mass matrix representation.
%
{\scriptsize
\begin{equation}
\label{15}
\begin{split}
M_u^\prime&=
\begin{pmatrix}
0&0&487.338\\
0&638&0\\
487.338 & 0&172099
\end{pmatrix}\textrm{MeV},
\\
M_d^\prime&=\begin{pmatrix}
 0 & 12.6421\,e^{-3.11099i}& 15.3817\,e^{-0.656498i}\\
12.6421e^{3.11099i} &-49.0183& 119.96\,e^{-0.000368804i}\\
15.3817\,e^{0.656498i} & 119.96\,e^{0.000368804i}& 2854.97
\end{pmatrix}\textrm{MeV}.
\end{split}
\end{equation}}
%
New  equivalent four-texture  zeros representations can be obtained using the former representation. For example, if we use  unitary  matrices looking like
%
\begin{equation}
\label{3-18}
U_1=\begin{pmatrix}	
                                 1&0&0\\
0&0&1\\
0&1&0
                                \end{pmatrix},
\end{equation}
and apply them
to~(\ref{15}), it allows us to obtain new non-parallel four-texture zeros  representations. For the case~(\ref{3-18}), we have
%
{\scriptsize
\begin{equation}
\label{16}
\begin{split}
M_u&=
\begin{pmatrix}
 0&487.338&0\\
487.338&172099&0\\
0 & 0&638
\end{pmatrix}\textrm{MeV},
\\
M_d&=\begin{pmatrix}
 0 & 15.3817\,e^{-0.656498i}& 12.6421\,e^{-3.11099i}\\
 15.3817\,e^{0.656498i}&2854.97 & 119.96\,e^{0.000368804i}\\
 12.6421e^{3.11099i}& 119.96\,e^{-0.000368804i}& -49.0183
\end{pmatrix}\textrm{MeV}.
\end{split}
\end{equation}}
%
where 
some of their entries have been permuted.

We have found typical non-parallel four-texture zeros quark mass matrix representations. The WB was applied by using  simple unitary matrices like~(\ref{9}). The process is more difficult if we want to find parallel  texture zeros in quark mass matrices.


\subsection{Parallel four-texture zeros }
\label{sIIIC}
Let us begin implementing a method that we shall apply later to special cases.
Let us start by giving the following structure for the up matrix elements
\footnote{It is sufficient to consider that the mass matrix be real and symmetric, since the phases may be included later by means of a WB process.}

\begin{equation}
\label{30}
 M_u=\begin{pmatrix}
      0&|C_u|&0\\
|C_u|&\tilde B_u&|B_u|\\
0&|B_u|&A_u
     \end{pmatrix},
\end{equation}
where $\tilde B_u$ and $A_u$ are real numbers.  
The mass matrix $M_u$ can be diagonalized using the transformation
\begin{equation}
\label{31}
 O_u^\dag M_uO_u=
\begin{pmatrix}
          \lambda_{1u}&&\\
&\lambda_{2u}&\\
&&\lambda_{3u}
         \end{pmatrix},
\end{equation}
where the exact analytical result of $O_u$ is~\cite{b4}
\begin{widetext}
\begin{equation} 
\label{32}
{
 O_u=\begin{pmatrix}
     e^{ix} \rho\sqrt{\frac{\lambda_{2u}\lambda_{3u}(A_u-\lambda_{1u})}{A_u(\lambda_{2u}-\lambda_{1u})(\lambda_{3u}-\lambda_{1u})}}&e^{iy}\eta
\sqrt{\frac{\lambda_{1u}\lambda_{3u}(\lambda_{2u}-A_u)}{A_u(\lambda_{2u}-\lambda_{1u})(\lambda_{3u}-\lambda_{2u})}}&
\sqrt{\frac{\lambda_{1u}\lambda_{2u}(A_u-\lambda_{3u})}{A_u(\lambda_{3u}-\lambda_{1u})(\lambda_{3u}-\lambda_{2u})}}\\
&&&\\[-2mm]
-e^{ix}\eta\sqrt{\frac{\lambda_{1u}(\lambda_{1u}-A_u)}{(\lambda_{2u}-\lambda_{1u})(\lambda_{3u}-\lambda_{1u})}}&
e^{iy}\sqrt{\frac{\lambda_{2u}(A_u-\lambda_{2u})}{(\lambda_{2u}-\lambda_{1u})(\lambda_{3u}-\lambda_{2u})}}&
\rho\sqrt{\frac{\lambda_{3u}(\lambda_{3u}-A_u)}{(\lambda_{3u}-\lambda_{1u})(\lambda_{3u}-\lambda_{2u})}}\\
&&&\\[-2mm]
e^{ix}\eta\sqrt{\frac{\lambda_{1u}(A_u-\lambda_{2u})(A_u-\lambda_{3u})}{A_u(\lambda_{2u}-\lambda_{1u})(\lambda_{3u}-\lambda_{1u})}}&
-e^{iy}\rho\sqrt{\frac{\lambda_{2u}(A_u-\lambda_{1u})(\lambda_{3u}-A_u)}{A_u(\lambda_{2u}-\lambda_{1u})(\lambda_{3u}-\lambda_{2u})}}&
\sqrt{\frac{\lambda_{3u}(A_u-\lambda_{1u})(A_u-\lambda_{2u})}{A_u(\lambda_{3u}-\lambda_{1u})(\lambda_{3u}-\lambda_{2u})}}
     \end{pmatrix},}
\end{equation}
\end{widetext}
where $\eta\equiv\lambda_{2u}/m_c=+1$ or $-1$ and $\rho\equiv\lambda_{3u}/m_t=+1$ or $-1$ corresponding to the possibility $(\lambda_{1u},\lambda_{2u},\lambda_{3u})=(-m_u,m_c,m_t)$, $(\lambda_{1u},\lambda_{2u},\lambda_{3u})=(m_u,-m_c,m_t)$ or $(\lambda_{1u},\lambda_{2u},\lambda_{3u})=(m_u,m_c,-m_t)$. The arbitrary phase factors 
 in~(\ref{32}) were included, in order that given them appropriated values, the generated  CKM matrix  becomes compatible with the chosen convention~(\ref{3.2})~\footnote{It is not necessary to include a phase factor in the third column of $O_u$, since we can factor out it.}.
Note that $\tilde B_u$, $|B_u|$ and $|C_u|$ can be expressed in terms of $\lambda_{iu}$~$(i=1,2, 3)$ and $A_u$, using invariant matrix functions as follows
%
\begin{align}
\label{3.18}
\text{tr} M_u&\Rightarrow \tilde B_u=\lambda_{1u}+\lambda_{2u}+\lambda_{3u}-A_u,\\
\label{34a}
\textrm{tr} M_u^2 &\Rightarrow|B_u|=\sqrt{\frac{(A_u-\lambda_{1u})(A_u-\lambda_{2u})(\lambda_{3u}-A_u)}{A_u}},\\
\label{35a}
\det M_u&\Rightarrow|C_u|=\sqrt{\frac{-\lambda_{1u}\lambda_{2u}\lambda_{3u}}{A_u}},
\end{align}
where ``tr'' and ``det'' are the trace and the determinant  respectively. 
The matrix $O_u$ can be seen as the unitary matrix such that the WB transformation
transforms the representation~(\ref{3}) into the form
%
\begin{align}
\label{41b}
 M_u^\prime&=
O_u\begin{pmatrix}
          \lambda_{1u}&&\\
&\lambda_{2u}&\\
&&\lambda_{3u}
         \end{pmatrix}O_u^\dag=\begin{pmatrix}
      0&|C_u|&0\\
|C_u|&\tilde B_u&|B_u|\\
0&|B_u|&A_u
     \end{pmatrix},
\\
\label{37}
 M_d^\prime&=O_u(VD_dV^\dag) O_u^\dag=
\begin{pmatrix}
 X_{(A_u,x,y)}&C_d&Y_{(A_u,x,y)}\\
C_d^*&\tilde B_d& B_d\\
Y^*_{(A_u,x,y)}& B_d^* & A_d
\end{pmatrix},
\end{align}
where the elements of $M_d^\prime$ depends on three parameters $A_u,x$ and $y$. To complete the analysis, we must obtain neglected values at the entries $(1,1)$ and $(1,3)$ compared with the remaining elements of the matrix $M^\prime_d$. Then we have to solve three equations
{\footnotesize
\begin{equation}
\label{38}
X_{(A_u,x,y)}=0,\:\textrm{Re}[Y_{(A_u,x,y)}]=0, \: \textrm{and} \:\: \textrm{Im}[Y_{(A_u,x,y)}]=0
\end{equation}}
\hspace{-1mm}where ``Re'' refers to the real part and ``Im'' the imaginary part of the function. 
In the process the following details must be taken into account:

%
\begin{itemize}
\item The formulas~(\ref{3.18}) through~(\ref{35a}) must be real numbers. Therefore, the   parameter $A_u$ is restricted to lie into an interval. Let us see the different possibilities
\begin{itemize}
 \item If $\lambda_{1u}=-m_u$, $\lambda_{2u}=m_c$ and $\lambda_{3u}=m_t$ then
\begin{equation}
\label{40}
 m_c<A_u<m_t.
\end{equation}
\item If $\lambda_{1u}=m_u$, $\lambda_{2u}=-m_c$ and $\lambda_{3u}=m_t$ then
\begin{equation}
\label{3.27}
 m_u<A_u<m_t.
\end{equation}
\item If $\lambda_{1u}=m_u$, $\lambda_{2u}=m_c$ and $\lambda_{3u}=-m_t$ then
\begin{equation}
 m_u<A_u<m_c.
\end{equation}
where the hierarchy~(\ref{19d}) was considered.
\end{itemize}
\item The phases given in~(\ref{32}) 
could have been included initially  in the transformation~(\ref{21b}), instead to write them explicitly in the matrix  $O_u$. The validity of this point of view is checked by observing that the matrix~(\ref{32}) can be decomposed as the product of two matrices, where the right hand side contains the phases as follows
\begin{equation}
 O_u=O_{u(x=0,y=0)}\:\begin{pmatrix}
e^{ix}&&\\
& e^{iy}&\\
&& 1
\end{pmatrix},
\end{equation}
such that, after replacing this decomposition into~(\ref{37}) and comparing with~(\ref{21b}), we conclude that both points of view concur.
\end{itemize}
In appendix~\ref{a2}, we will work a case previously studied in the paper~\cite{b0} and replicate the results presented there by using the techniques implemented here.


\subsubsection{Example 1: parallel four-texture zeros}
We are mainly concerned to find four-texture zeros with the recent data given in Section~\ref{A}. Let us take the following case
\begin{align}
 \lambda_{1u}&=-m_u,\lambda_{2u}=m_c,\lambda_{3u}=m_t,\\
 \lambda_{1d}&=-m_d,\lambda_{2d}=m_s,\lambda_{3d}=m_b.
\end{align}
We have, in the u-diagonal representation,  the following mass matrix representation.
{
{\scriptsize
\begin{equation}
\label{3.45}
\begin{split}
M_u&=\begin{pmatrix}
      -1.38&0&0\\
0&638&0\\
0&0&172100
     \end{pmatrix}\text{MeV},
\\
 M_d&=VD_dV^\dag\\
&=\begin{pmatrix}
      0.253114& 13.2691 - 0.386919 i & 3.01706 - 9.38676 i\\
13.2691 + 0.386919 i& 58.7203 & 115.45 + 0.043146 i\\
3.01706 + 9.38676 i & 115.45 - 0.043146 i &2855.21
     \end{pmatrix}\text{MeV}.
\end{split}
\end{equation}}}
\hspace{-1mm}Making a WB transformation on~(\ref{3.45}), using the unitary matrix $O_u$~(Eq. \ref{32}),
the following conditions
\begin{equation}
\label{65}
\begin{split}
 M_{d(1,1)}^\prime(A_u,x_1,x_2,y_1,y_2)&=0,
\\
\textrm{Re}\left[ M_{d(1,3)}^\prime(A_u,x_1,x_2,y_1,y_2)\right]&=0,
\\
 \textrm{Im} \left[M_{d(1,3)}^\prime(A_u,x_1,x_2,y_1,y_2)\right]&=0,
\end{split}
\end{equation}
are established, in order to find zero entries in (1,1), (1,3) and (3,1) of the resulting matrix $M_d^\prime=O_uM_dO_u^\dag$; where the  phases given in $O_u$ has been defined as $e^{ix}=\cos x+i\sin x=x_1+ix_2$ and $e^{iy}=\cos y+i\sin y=y_1+iy_2$, such that
\begin{equation}
\label{66}
 x_1^2+x_2^2=1\quad \textrm{and}\quad  y_1^2+y_2^2=1.
\end{equation}
%
Eqs.~(\ref{65}) and~(\ref{66}) gives the following exact solution.
{
\begin{equation}
\begin{split}
A_u&= 153231~\textrm{MeV},\: x_1 =0.883194, \: x_2= -0.469007,\\
 y_1&=0.202996,\quad y_2=0.97918.
\end{split}
\end{equation}}
%
Finally, we obtain an exact parallel  four-texture zeros mass matrix representation.
{\scriptsize
\begin{align}
 M_u^\prime&=O_uM_uO_u^\dag=
\begin{pmatrix}
 0 & 31.4461 & 0\\
31.4461 & 19505.7 & 53659.2\\
0 & 53659.2 & 153231
\end{pmatrix}\textrm{MeV},\\
\begin{split}
M_d^\prime&=O_uM_dO_u^\dag,\\
&=
\begin{pmatrix}
 0 & -1.43578 - 13.3956 i & 0\\
-1.43578 + 13.3956 i & 381.367 & 893.365+ 113.383 i \\
0 &  893.365 - 113.383 i & 2532.81
\end{pmatrix}\text{MeV},
\end{split}
\end{align}}
In the same way, we can find other non-equivalent parallel four-texture zeros representations. Let us look  another case.


\subsubsection{Example 2: another parallel four-texture zeros model}
Another possibility that works well is
\begin{align}
 \lambda_{1u}&=m_u,\lambda_{2u}=m_c,\lambda_{3u}=-m_t,\\
 \lambda_{1d}&=m_d,\lambda_{2d}=m_s,\lambda_{3d}=-m_b,
\end{align}
from which, we have $A_u= 7.34102~\textrm{MeV}, x_1=0.998393,$ $x_2=-0.0566637, y_1 =0.999664$
and $y_2=0.0259074$. Thus, the corresponding parallel four-texture zeros mass matrix representation is 
{\footnotesize
\begin{align}
 M_u^\prime&=O_uM_uO_u^\dag=
\begin{pmatrix}
 0 & 4543.2& 0\\
4543.2 & -171468. & 9388.13\\
0 &9388.13& 7.34102
\end{pmatrix}\textrm{MeV},\\
\begin{split}
M_d^\prime&=O_uM_dO_u^\dag\\
&=
\begin{pmatrix}
0 & 123.93 + 10.0184 i & 0\\
123.93 - 10.0184 i& -2829.92 & 267.035 + 1.39152 i \\
0&  267.035 - 1.39152 i & 29.738
\end{pmatrix}\textrm{MeV}.
\end{split}	
\end{align}}

\section{Numerical Five-Texture Zeros}
\label{sIV}
Now, let us try to find five-texture zeros for the quark mass matrix sector. If this cannot be achieved, we can conclude that five and six-texture zeros are not viable models. For that, we will use the mathematical tools previously implemented in Sect.~\ref{sIIIC}. We shall begin as usual by proposing a texture zeros configuration, 
in this case with three zeros for the up/down quark mass matrix\footnote{A model with four zeros in the up/down quark mass matrix is not realistic.}, and see how many zeros can be reached for the down/up quark mass matrix. In principle, there are many possibilities, but many of them are equivalent ones. In total, there are two non-equivalent cases, depending on the number of zeros included in their diagonal entries. Therefore, we have only two possibilities: one-zero  or  two-zero in diagonal entries. Let us name them as {\it one-zero family} and {\it two-zero family}, respectively. With an appropriated unitary matrix and performing the corresponding WB transformation the other possibilities are obtained. In the Table~\ref{ta1}  both families are indicated, which summarizes the equivalent possibilities for each case. Let us study each family.


%
\begin{table}[htb]
{\footnotesize
\begin{tabular}{|p{2.09cm}|p{3.17cm}|p{3.02cm}|}
\hline
\hline
\multicolumn{1}{|c|}{}& \multicolumn{1}{c|}{}&\multicolumn{1}{c|}{}\\[-2mm]
\textbf{Unitary matrix}&\textbf{Two-zero Family}  $(p_i\:M_{u,d}\:p_i^T)$ & \textbf{One-zero family} $(p_i\:M_{u,d}\:	p_i^T)$\\[1mm]
\hline\hline
\multicolumn{1}{|c|}{}& \multicolumn{1}{c|}{}&\multicolumn{1}{c|}{}\\[-1mm]
$p_1=\begin{pmatrix}
      1&&\\
&1&\\
&&1
     \end{pmatrix}
$&$\begin{pmatrix}
      0& |C_{u,d}|& 0\\
|C_{u,d}|& 0& |B_{u,d}|\\
0& |B_{u,d}|& A_{u,d}
     \end{pmatrix}$&
$\begin{pmatrix}
      0& |B_{u,d}|& 0\\
|B_{u,d}|& C_{u,d}& 0\\
0& 0& A_{u,d}
     \end{pmatrix}$\\[5mm]
\hline
\multicolumn{1}{|c|}{}& \multicolumn{1}{c|}{}&\multicolumn{1}{c|}{}\\[-1mm]
$p_2=\begin{pmatrix}
      1&&\\
&&1\\
&1&
     \end{pmatrix}
$&$\begin{pmatrix}
      0&0 & |C_{u,d}|\\
0& A_{u,d}& |B_{u,d}|\\
|C_{u,d}|& |B_{u,d}|& 0
     \end{pmatrix}$&
$\begin{pmatrix}
 0& 0& |B_{u,d}|\\
0& A_{u,d}& 0\\
|B_{u,d}|& 0& C_{u,d}
     \end{pmatrix}$\\[5mm]
\hline
\multicolumn{1}{|c|}{}& \multicolumn{1}{c|}{}&\multicolumn{1}{c|}{}\\[-1mm]
$p_3=\begin{pmatrix}
     &&1 \\
&1& \\
1&& 
     \end{pmatrix}
$&$\begin{pmatrix}
      A_{u,d}& |B_{u,d}|& 0\\
|B_{u,d}|& 0&|C_{u,d}| \\
0& |C_{u,d}|& 0
     \end{pmatrix}$&
$\begin{pmatrix}
      A_{u,d}& 0& 0\\
0& C_{u,d}& |B_{u,d}|\\
0& |B_{u,d}|& 0
     \end{pmatrix}$\\[5mm]
\hline
\multicolumn{1}{|c|}{}& \multicolumn{1}{c|}{}&\multicolumn{1}{c|}{}\\[-1mm]
$p_4=\begin{pmatrix}
     &1& \\
1&& \\
 &&1
     \end{pmatrix}
$&$\begin{pmatrix}
      0& |C_{u,d}|&|B_{u,d}|\\
|C_{u,d}|& 0& 0\\
|B_{u,d}|& 0& A_{u,d}
     \end{pmatrix}$&
$\begin{pmatrix}
      |C_{u,d}|& |B_{u,d}|& 0\\
|B_{u,d}|& 0& 0\\
0& 0& A_{u,d}
     \end{pmatrix}$\\[5mm]
\hline
\multicolumn{1}{|c|}{}& \multicolumn{1}{c|}{}&\multicolumn{1}{c|}{}\\[-1mm]
$p_5=\begin{pmatrix}
      &&1 \\
1&& \\
&1&
     \end{pmatrix}
$&$\begin{pmatrix}
      A_{u,d}& 0& |B_{u,d}|\\
0& 0& |C_{u,d}|\\
|B_{u,d}|& |C_{u,d}|& 0
     \end{pmatrix}$&
$\begin{pmatrix}
      A_{u,d}& 0& 0\\
0& 0& |B_{u,d}|\\
0& |B_{u,d}|& C_{u,d}
     \end{pmatrix}$\\[5mm]
\hline
\multicolumn{1}{|c|}{}& \multicolumn{1}{c|}{}&\multicolumn{1}{c|}{}\\[-1mm]
$p_6=\begin{pmatrix}
     &1& \\
&&1\\
1&& 
     \end{pmatrix}
$&$\begin{pmatrix}
      0&  |B_{u,d}|&|C_{u,d}|\\
 |B_{u,d}|&A_{u,d}&0 \\
|C_{u,d}|& 0& 0
     \end{pmatrix}$&
$\begin{pmatrix}
      C_{u,d}& 0& |B_{u,d}|\\
0& A_{u,d}& 0\\
|B_{u,d}|& 0& 0
     \end{pmatrix}$\\[5mm]
\hline
\end{tabular}
}
\vspace{-3mm}
\caption{One and two-zero Family.}
\label{ta1}
\end{table}
%

\subsection{Two-zero family}
In what follows, we work the \textit{u-diagonal } and \textit{d-diagonal } cases simultaneously.
The standard representation for the two-zero family is 
\begin{equation}
\label{77}
 M_{u,d}=\begin{pmatrix}
      0& |C_{u,d}|& 0\\
|C_{u,d}|& 0& |B_{u,d}|\\
0& |B_{u,d}|& A_{u,d}
     \end{pmatrix}.
\end{equation}
 and its diagonalization matrix satisfies the following relation  
\begin{equation}
\label{78a}
 O_{u,d}^\dag M_{u,d}O_{u,d}=\begin{pmatrix}
              \lambda_{1{u,d}}&&\\
&\lambda_{2{u,d}}&\\
&&\lambda_{3{u,d}}
             \end{pmatrix},
\end{equation}
where one and only one $\lambda_{i{u,d}}$ is assumed to be a negative number.
The invariant quantities ``$\det$'' and ``trace'' applied on~(\ref{77}) and~(\ref{78a}) 
\begin{align}
 \textrm{tr}M_{u,d}&=A_{u,d}=\lambda_{1{u,d}}+\lambda_{2{u,d}}+\lambda_{3{u,d}},\\
\begin{split}
\textrm{tr}M_{u,d}^2&=A_{u,d}^2+2|B_{u,d}|^2+2|C_{u,d}|^2\\
&=\lambda_{1{u,d}}^2+\lambda_{2{u,d}}^2+\lambda_{3{u,d}}^2,
\end{split}
\\
\textrm{det}M_{u,d}&=-A_{u,d}|C_{u,d}|^2=\lambda_{1{u,d}}\lambda_{2{u,d}}\lambda_{3{u,d}},
\end{align}
allow us to express the parameters of~(\ref{77}) in terms of its eigenvalues 
\begin{align}
 A_{u,d}&=\lambda_{1{u,d}}+\lambda_{2{u,d}}+\lambda_{3{u,d}},\\
\label{84}
|B_{u,d}|&=\sqrt{-\frac{(\lambda_{1{u,d}}+\lambda_{2{u,d}})(\lambda_{1{u,d}}+\lambda_{3{u,d}})(\lambda_{2{u,d}}+\lambda_{3{u,d}})}{A_{u,d}}},\\
\label{85}
|C_{u,d}|&=\sqrt{-\frac{\lambda_{1{u,d}}\lambda_{2{u,d}}\lambda_{3{u,d}}}{A_{u,d}}}.
\end{align}
From expression~(\ref{85}), together with~(\ref{2.9}), we have that
\begin{equation}
\label{4.9}
 A_{u,d}>0,
\end{equation}
and using~(\ref{84}) and the hierarchy~(\ref{19d}) we found that only one possibility is permitted
\begin{equation}
\label{4.10}
 \lambda_{1{u,d}},\lambda_{3{u,d}}>0\quad\textrm{and}\quad\lambda_{2{u,d}}<0.
\end{equation}
For the u-diagonal case, the diagonalization matrix~(\ref{32}) becomes
{
\begin{equation}
 O_u=\begin{pmatrix}
      0.99892 e^{ix}& -0.0464583 e^{iy}&0.0000104863\\
0.0463719e^{ix}& 0.997078 e^{iy}& 0.0607083\\
-0.00283086e^{ix}& -0.0606422e^{iy}& 0.998156
     \end{pmatrix},
\end{equation}
and for the d-diagonal case, the diagonalization matrix is given by
\begin{equation}
O_d=\begin{pmatrix}
  0.980856e^{ix}& -0.194731e^{iy}&0.000682127\\
0.19251e^{ix}&0.970182 e^{iy}&0.147267\\
-0.0293392e^{ix}& -0.144316 e^{iy}&0.989097
 \end{pmatrix}.
\end{equation}}
As you can see, in both cases, we are treating with quasi diagonal matrices.

Performing the WB transformation using the unitary matrix $O_{u,d}$ we have
%
\begin{align}
\begin{split}
M_{u,d}^{\prime}&=O_{u,d}\begin{pmatrix}
              \lambda_{1{u,d}}&&\\
&\lambda_{2{u,d}}&\\
&&\lambda_{3{u,d}}
             \end{pmatrix}O_{u,d}^\dag\:,
\end{split}\\
&=\begin{pmatrix}
      0& |C_{u,d}|& 0\\
|C_{u,d}|& 0& |B_{u,d}|\\
0& |B_{u,d}|& A_{u,d}
     \end{pmatrix}\quad \textrm{and}\quad \\
\label{91a}
M_{d,u}^\prime&=O_{d,u}M_{d,u}O_{d,u}^\dag,
\end{align}
where the matrices
\begin{equation}
M_d=VD_dV^{\dag}\quad\textrm{and}\quad M_u=V^\dag D_uV,
\end{equation}
depend on if we work with either \textit{the u-diagonal} or \textit{the d-diagonal  case}. 

In order to facilitate the calculus we define the following new variables
\begin{align}
\label{89}
\begin{split}
 e^{ix}&=
x_1+ix_2, \: \textrm{with}\quad x_1^2+x_2^2=1,
\\
e^{iy}&=
y_1+iy_2, \:\textrm{with}\quad y_1^2+y_2^2=1, 
\end{split}
\end{align}
where their norms satisfy
\begin{equation}
\label{4.18}
 |x_1|,|x_2|\leq1,
\quad\textrm{and}\quad |y_1|,|y_2|\leq1.
\end{equation}
With the former definitions, the elements of the matrix $M_{d,u}^\prime$ defined in~(\ref{91a}) have now a polynomial form in each case considered: $\lambda_{1d}=-m_d$ or $\lambda_{2d}=-m_s$ or  $\lambda_{3d}=-m_b$ for the u-digonal case~(or $\lambda_{1u}=-m_d$ or $\lambda_{2u}=-m_s$ or  $\lambda_{3u}=-m_b$ for the d-diagonal case) .
The results are summarized in Tables~(\ref{t2}) and~(\ref{t3})
%
\begin{widetext}
\
\begin{table}[htb]
\centering
\begin{ruledtabular}
\begin{tabular}{|c|p{5cm}|p{5cm}|p{5cm}|}
\multicolumn{1}{|c|}{}& \multicolumn{1}{c}{}&\multicolumn{1}{c}{}&\multicolumn{1}{c|}{}\\[-3mm]
 $M_d^\prime$&\multicolumn{3}{c|}{\textrm{Negative mass eigenvalue}
}\\[1mm]
\hline
\multicolumn{1}{|c|}{}& \multicolumn{1}{c|}{}&\multicolumn{1}{c|}{}&\multicolumn{1}{c|}{}\\[-2mm]
{\footnotesize entries}&Case 1. $\lambda_{1d}=-m_d$\:(MeV)&Case 2. $\lambda_{2d}=-m_s$\:(MeV)&Case 3. $\lambda_{3d}=-m_b$\:(MeV)\\[1mm] \hline\hline
{\footnotesize$M_d^\prime(1,1)$}&{\footnotesize$0.758616 + 0.0000632072x_1 + 0.000196652 x_2 - 
  0.000112489 y_1 - 1.23159 x_1 y_1 - 0.0359124 x_2 y_1 + 0.0359124 x_1 y_2 - 
  1.23159 x_2 y_2$} 
  & {\footnotesize$-0.575839 + 0.0000858823 x_1+0.000196682 x_2 - 0.000116848 y_1 + 1.20436 x_1 y_1 - 0.0359178 x_2 y_1+0.0359178 x_1 y_2 + 1.20436 x_2 y_2$} &{\footnotesize$11.261 - 0.0000849534 x_1 - 0.00019705 x_2 + 0.000116858 y_1 - 
  1.0895 x_1 y_1 + 0.0359851 x_2 y_1 - 0.0359851 x_1 y_2 - 1.0895 x_2 y_2$}\\[1mm]
\hline
{\footnotesize$M_d^\prime(1,2)$}&{\footnotesize$-5.41488 + 0.182964 x_1 + 0.569243 x_2 - 0.324408 y_1 + 13.1875 x_1 y_1 + 
  0.384538 x_2 y_1 + 0.000121238 y_2 - 0.384538 x_1 y_2 + 13.1875 x_2 y_2+i(-0.569234 x_1 + 0.182961 x_2 - 0.00012214 y_1 - 0.386206 x_1 y_1 + 13.2447 x_2 y_1 - 
  0.326823 y_2 - 13.2447 x_1 y_2 - 0.386206 x_2 y_2)$}&{\footnotesize$4.52621 + 0.248601 x_1 + 0.56933 x_2 - 0.336979 y_1 - 
  12.8959 x_1 y_1 + 0.384597 x_2 y_1 - 0.000121238 y_2 - 0.384597 x_1 y_2 - 
  12.8959 x_2 y_2+i(-0.569321 x_1 + 0.248597 x_2 + 0.00012214 y_1 - 0.386264 x_1 y_1 - 12.9519 x_2 y_1 - 
  0.339487 y_2 + 12.9519 x_1 y_2 - 0.386264 x_2 y_2)$}&{\footnotesize$-4.05674 - 0.245913 x_1 - 0.570396 x_2 + 0.337008 y_1 + 11.6661 x_1 y_1 - 
  0.385317 x_2 y_1 + 0.000109807 y_2 + 0.385317 x_1 y_2 + 11.6661 x_2 y_2+i(0.570386 x_1 - 0.245909 x_2 - 0.000110625 y_1 + 0.386987 x_1 y_1 + 11.7166 x_2 y_1 + 
  0.339516 y_2 - 11.7166 x_1 y_2 + 0.386987 x_2 y_2)$}\\[1mm]
\hline
{\footnotesize$M_d^\prime(1,3)$}&{\footnotesize$0.359323 + 3.00824 x_1 + 9.35933 x_2 - 5.35379 y_1 - 
  0.802056 x_1 y_1 - 0.0233874 x_2 y_1 + 0.00200082 y_2 + 0.0233874 x_1 y_2 - 
  0.802056 x_2 y_2+i(-9.35933 x_1 + 3.00824 x_2 - 0.00200077 y_1 + 0.0234892 x_1 y_1 - 0.805546 x_2 y_1 - 
  5.35364 y_2 + 0.805546 x_1 y_2 + 0.0234892 x_2 y_2)$ }&{\footnotesize$ -0.245286 + 4.08743 x_1 + 9.36075 x_2 - 5.56125 y_1 + 0.784325 x_1 y_1 - 
  0.023391 x_2 y_1 - 0.00200082 y_2 + 0.023391 x_1 y_2 + 0.784325 x_2 y_2+i(-9.36075 x_1 + 4.08743 x_2 + 0.00200077 y_1 + 0.0234928 x_1 y_1 + 0.787738 x_2 y_1 - 
  5.56109 y_2 - 0.787738 x_1 y_2 + 0.0234928 x_2 y_2)$ }&{\footnotesize$ 0.216621 - 4.04322 x_1 - 9.37828 x_2 + 5.56172 y_1 - 
  0.709524 x_1 y_1 + 0.0234348 x_2 y_1 + 0.00181218 y_2 - 0.0234348 x_1 y_2 - 
  0.709524 x_2 y_2+i(9.37828 x_1 - 4.04322 x_2 - 0.00181213 y_1 - 0.0235368 x_1 y_1 - 0.712611 x_2 y_1 + 
  5.56157 y_2 + 0.712611 x_1 y_2 - 0.0235368 x_2 y_2)$ }\\[1mm]
\hline
{\footnotesize$M_d^\prime(2,2)$}&{\footnotesize$127.279 + 0.016987 x_1 + 0.0528505 x_2 + 13.9766 y_1 + 
  1.22703 x_1 y_1 + 0.0357795 x_2 y_1 - 0.00522333 y_2 - 0.0357795 x_1 y_2 + 
  1.22703 x_2 y_2$}&{\footnotesize$-86.9431 + 0.023081 x_1 + 0.0528585 x_2 + 14.5182 y_1 - 1.19991 x_1 y_1 + 
  0.035785 x_2 y_1 + 0.00522333 y_2 - 0.035785 x_1 y_2 - 1.19991 x_2 y_2$}&{\footnotesize$87.5349 - 0.0228314 x_1 - 0.0529575 x_2 - 14.5194 y_1 + 
  1.08547 x_1 y_1 - 0.035852 x_2 y_1 - 0.00473086 y_2 + 0.035852 x_1 y_2 + 
  1.08547 x_2 y_2$}\\[1mm]
\hline
{\footnotesize$M_d^\prime(2,3)$}&{\footnotesize$165.914 + 0.13913 x_1 + 0.432866 x_2 + 114.475 y_1 - 
  0.0747673 x_1 y_1 - 0.00218017 x_2 y_1 - 0.0427818 y_2 + 0.00218017 x_1 y_2 - 
  0.0747673 x_2 y_2+i(-0.436093 x_1 + 0.140167 x_2 + 0.0430994 y_1 + 0.000139144 x_2 y_1 + 115.325 y_2 - 
  0.000139144 x_1 y_2) $}&{\footnotesize$178.932 + 0.189042 x_1 + 0.432932 x_2 + 118.911 y_1 + 
  0.0731144 x_1 y_1 - 0.0021805 x_2 y_1 + 0.0427818 y_2 + 0.0021805 x_1 y_2 + 
  0.0731144 x_2 y_2+i(-0.436159 x_1 + 0.190451 x_2 - 0.0430994 y_1 - 0.000136068 x_2 y_1 + 119.794 y_2 + 
  0.000136068 x_1 y_2)$}&{\footnotesize$-178.968 - 0.186998 x_1 - 0.433742 x_2 - 118.921 y_1 - 0.0661415 x_1 y_1 + 
  0.00218458 x_2 y_1 - 0.0387482 y_2 - 0.00218458 x_1 y_2 - 0.0661415 x_2 y_2+i(0.436975 x_1 - 0.188392 x_2 + 0.0390359 y_1 + 0.000123091 x_2 y_1 - 119.804 y_2 - 
  0.000123091 x_1 y_2)$}\\[1mm]
\hline
{\footnotesize$M_d^\prime(3,3)$}&{\footnotesize$2845.12- 0.0170502 x_1 - 0.0530471 x_2 - 13.9765 y_1 + 
  0.00455581 x_1 y_1 + 0.000132844 x_2 y_1 + 0.00522329 y_2 - 0.000132844 x_1 y_2 + 
  0.00455581 x_2 y_2$}&{\footnotesize$2844.14- 0.0231669 x_1 - 0.0530552 x_2 - 14.5181 y_1 - 
  0.00445509 x_1 y_1 + 0.000132865 x_2 y_1 - 0.00522329 y_2 - 0.000132865 x_1 y_2 - 
  0.00445509 x_2 y_2$}&{\footnotesize$-2844.14 + 0.0229163 x_1 + 0.0531545 x_2 + 14.5193 y_1 + 0.00403021 x_1 y_1 - 
  0.000133113 x_2 y_1 + 0.00473083 y_2 + 0.000133113 x_1 y_2 + 0.00403021 x_2 y_2$}\\[1mm]
%
\end{tabular}
\end{ruledtabular}
\vspace{-4mm}
\caption{\textit{The u-diagonal representation}: the ``down'' mass matrix entries  for the two-zero family case.}
\label{t2}
\end{table} 

\begin{table}[htb]
\centering
\begin{ruledtabular}
\begin{tabular}{|c|p{5cm}|p{5cm}|p{5.15cm}|}
\multicolumn{1}{|c|}{}& \multicolumn{1}{c}{}&\multicolumn{1}{c}{}&\multicolumn{1}{c|}{}\\[-3mm]
 {\footnotesize $M_u^\prime$}&\multicolumn{3}{c|}{\textrm{Negative mass eigenvalue}}\\[1mm]
\hline
\multicolumn{1}{|c|}{}& \multicolumn{1}{c|}{}&\multicolumn{1}{c|}{}&\multicolumn{1}{c|}{}\\[-3mm]
{\footnotesize entries}&Case 1. $\lambda_{1u}=-m_u$\:(MeV)&Case 2. $\lambda_{2u}=-m_c$\:(MeV)&Case 3. $\lambda_{3u}=-m_t$\:(MeV)\\[1mm] \hline\hline
{\footnotesize$M_u^\prime(1,1)$}&{\footnotesize $151.93+ 1.84869 x_1 - 0.735842 x_2 + 1.839 y_1 + 
  74.8244 x_1 y_1 - 8.85442 x_2 y_1 + 0.0337901 y_2 + 8.85442 x_1 y_2 + 74.8244 x_2 y_2$} &{\footnotesize $-59.245 + 1.86453 x_1 - 0.735821 x_2 + 1.85259 y_1 - 32.2673 x_1 y_1 - 
  8.92032 x_2 y_1 + 0.0337892 y_2 + 8.92032 x_1 y_2 - 32.2673 x_2 y_2$}&{\footnotesize $64.2966- 1.86453 x_1 + 0.735833 x_2 - 1.85259 y_1 + 
  32.0358 x_1 y_1 + 8.92032 x_2 y_1 - 0.0337897 y_2 - 8.92032 x_1 y_2 + 32.0358 x_2 y_2$}\\[1mm]
\hline
{\footnotesize$M_u^\prime(1,2)$}&{\footnotesize $-300.727 + 199.742 x_1 - 79.504 x_2 + 193.933 y_1 - 179.051 x_1 y_1 + 
  21.1882 x_2 y_1 + 3.56336 y_2 - 21.1882 x_1 y_2 - 179.051 x_2 y_2+i(79.3596 x_1 + 199.379 x_2 - 3.73171 y_1 - 22.926 x_1 y_1 - 193.736 x_2 y_1 + 
  203.095 y_2 + 193.736 x_1 y_2 - 22.926 x_2 y_2)$}&{\footnotesize $132.633 + 201.453 x_1 - 79.5017 x_2 + 195.366 y_1 + 
  77.2138 x_1 y_1 + 21.3458 x_2 y_1 + 3.56326 y_2 - 21.3458 x_1 y_2 + 77.2138 x_2 y_2+i(79.3573 x_1 + 201.087 x_2 - 3.7316 y_1 - 23.0966 x_1 y_1 + 83.5468 x_2 y_1 + 
  204.596 y_2 - 83.5468 x_1 y_2 - 23.0966 x_2 y_2)$}&{\footnotesize $-131.697 - 201.453 x_1 + 79.503 x_2 - 195.366 y_1 - 76.6599 x_1 y_1 - 
  21.3458 x_2 y_1 - 3.56332 y_2 + 21.3458 x_1 y_2 - 76.6599 x_2 y_2+i(-79.3586 x_1 - 201.087 x_2 + 3.73166 y_1 + 23.0966 x_1 y_1 - 82.9474 x_2 y_1 - 
  204.596 y_2 + 82.9474 x_1 y_2 + 23.0966 x_2 y_2)$}\\[1mm]
\hline
{\footnotesize$M_u^\prime(1,3)$}&{\footnotesize $163.157 + 1340.29 x_1 - 533.481 x_2 + 1333.97 y_1 + 
  26.6073 x_1 y_1 - 3.1486 x_2 y_1 + 24.5107 y_2 + 3.1486 x_1 y_2 + 26.6073 x_2 y_2+i(533.503 x_1 + 1340.34 x_2 - 24.4856 y_1 + 3.41346 x_1 y_1 + 28.8455 x_2 y_1 + 
  1332.61 y_2 - 28.8455 x_1 y_2 + 3.41346 x_2 y_2)$}&{\footnotesize $98.7777 + 1351.77 x_1 - 533.466 x_2 + 1343.83 y_1 - 
  11.4741 x_1 y_1 - 3.17203 x_2 y_1 + 24.51 y_2 + 3.17203 x_1 y_2 - 11.4741 x_2 y_2+i(533.488 x_1 + 1351.83 x_2 - 24.4849 y_1 + 3.43886 x_1 y_1 - 12.4393 x_2 y_1 + 
  1342.46 y_2 + 12.4393 x_1 y_2 + 3.43886 x_2 y_2)$}&{\footnotesize $-98.9206 - 1351.77 x_1 + 533.475 x_2 - 1343.83 y_1 + 11.3918 x_1 y_1 + 
  3.17203 x_2 y_1 - 24.5103 y_2 - 3.17203 x_1 y_2 + 11.3918 x_2 y_2+i(-533.497 x_1 - 1351.83 x_2 + 24.4853 y_1 - 3.43886 x_1 y_1 + 12.3501 x_2 y_1 - 
  1342.46 y_2 - 12.3501 x_1 y_2 - 3.43886 x_2 y_2)$}\\[1mm]
\hline
{\footnotesize$M_u^\prime(2,2)$}&{\footnotesize $5396.4 + 78.3341 x_1 - 31.1797 x_2 - 1978.06 y_1 - 
  73.1657 x_1 y_1 + 8.65814 x_2 y_1 - 36.3452 y_2 - 8.65814 x_1 y_2 - 73.1657 x_2 y_2$}&{\footnotesize $3115.84 + 79.0053 x_1 - 31.1788 x_2 - 1992.67 y_1 + 
  31.552 x_1 y_1 + 8.72257 x_2 y_1 - 36.3441 y_2 - 8.72257 x_1 y_2 + 31.552 x_2 y_2$}&{\footnotesize $-3115.38 - 79.0051 x_1 + 31.1793 x_2 + 1992.67 y_1 - 31.3256 x_1 y_1 - 
  8.72257 x_2 y_1 + 36.3447 y_2 + 8.72257 x_1 y_2 - 31.3256 x_2 y_2$}\\[1mm]
\hline
{\footnotesize$M_u^\prime(2,3)$}&{\footnotesize $24777.1 + 257.091 x_1 - 102.331 x_2 - 6495.55 y_1 + 
  11.0171 x_1 y_1 - 1.30373 x_2 y_1 - 119.35 y_2 + 1.30373 x_1 y_2 + 11.0171 x_2 y_2+i(107.083 x_1 + 269.029 x_2 + 124.757 y_1 - 0.0158108 x_1 y_1 - 0.13361 x_2 y_1 - 
  6789.79 y_2 + 0.13361 x_1 y_2 - 0.0158108 x_2 y_2)$}&{\footnotesize $25116.1 + 259.294 x_1 - 102.328 x_2 - 6543.55 y_1 - 
  4.75103 x_1 y_1 - 1.31343 x_2 y_1 - 119.347 y_2 + 1.31343 x_1 y_2 - 4.75103 x_2 y_2+i(107.08 x_1 + 271.334 x_2 + 124.753 y_1 - 0.0159285 x_1 y_1 + 0.0576178 x_2 y_1 - 
  6839.96 y_2 - 0.0576178 x_1 y_2 - 0.0159285 x_2 y_2)$}&{\footnotesize $-25116.1 - 259.293 x_1 + 102.33 x_2 + 6543.55 y_1 + 4.71695 x_1 y_1 + 
  1.31343 x_2 y_1 + 119.349 y_2 - 1.31343 x_1 y_2 + 4.71695 x_2 y_2+i(-107.081 x_1 - 271.334 x_2 - 124.755 y_1 + 0.0159285 x_1 y_1 - 0.0572044 x_2 y_1 + 
  6839.96 y_2 + 0.0572044 x_1 y_2 + 0.0159285 x_2 y_2)$}\\[1mm]
\hline
{\footnotesize$M_u^\prime(3,3)$}&{\footnotesize $168118 - 80.1828 x_1 + 31.9155 x_2 + 1976.22 y_1 - 
  1.6587 x_1 y_1 + 0.196283 x_2 y_1 + 36.3114 y_2 - 0.196283 x_1 y_2 - 1.6587 x_2 y_2$}&{\footnotesize $168065. - 80.8698 x_1 + 31.9146 x_2 + 1990.82 y_1 + 
  0.715295 x_1 y_1 + 0.197744 x_2 y_1 + 36.3103 y_2 - 0.197744 x_1 y_2 + 
  0.715295 x_2 y_2$}&{\footnotesize $-168065. + 80.8696 x_1 - 31.9151 x_2 - 1990.82 y_1 - 0.710164 x_1 y_1 - 
  0.197744 x_2 y_1 - 36.3109 y_2 + 0.197744 x_1 y_2 - 0.710164 x_2 y_2$}\\[1mm]
\end{tabular}
\end{ruledtabular}
\vspace{-4mm}
\caption{\textit{The d-diagonal representation:} the ``up'' mass matrix entries for the two-zero family case.}
\label{t3}
\end{table}
\end{widetext}


\subsubsection{Analysis of ``down'' mass matrix.}
Table~(\ref{t2}) summarizes the components of  $M_d^\prime$ for \textit{the u-diagonal  case}.
By simple inspection, using~(\ref{4.18}), shows that is not possible to find zeros at  entries (2,2), (2,3) and~(3,3). And not solutions were found for either                                                                                                                                                              %
\begin{equation*}
\begin{split}
 {\rm Re}&[M_d^\prime(1,2)]=0,\quad{\rm Im}[M_d^\prime(1,2)]=0,\quad{\rm or}\\
{\rm Re}&[M_d^\prime(1,3)]=0,\quad{\rm Im}[M_d^\prime(1,3)]=0,
\end{split}
\end{equation*}
equations. Therefore, it is impossible to find two texture zeros into the down quark mass matrix coming from an u-diagonal representation for the two-zero family case.

\subsubsection{Analysis of ``up'' mass matrix  and a model with five-texture zeros.}
\label{s4a2}
Let us consider \textit{the d-diagonal  case.} The entries of matrix  $M_u^\prime$, after the WB transformation is made, are given in the Table~(\ref{t3}). According to the Table, only entries (1,2) and (1,3) deserve some attention. From which, only the cases $\lambda_{1u}=-m_u$ and $\lambda_{2u}=-m_c$ give an acceptable solution. 

For the first case, with $\lambda_{1u}=-m_u$, we have
\begin{align}
 M_u^\prime(1,2)&=0,\\
M_u^\prime(1,1)&\approx0,
\end{align}
where
{
\begin{equation}
\begin{split}
 x_1 &= 0.706984,\:\: y_1 =-0.540778,
\\ 
x_2 &= 0.70723,\: \:y_2 = -0.841165.
\end{split}
\end{equation}}
The corresponding five-texture zeros representation obtained, is.
{
{
\begin{subequations}
\label{4.22}
\begin{align}
 M_u^{\prime}&=\begin{pmatrix}
      0&0&-92.3618 + 157.694 i\\
0& 5748.17& 28555.1+ 5911.83 i\\
-92.3618 - 157.694 i& 28555.1 - 5911.83 i& 166988
     \end{pmatrix}\text{MeV},\\
M_d^{\prime}&=\begin{pmatrix}
     0& 13.9899& 0\\
13.9899 & 0&424.808\\
0& 424.808& 2796.9
    \end{pmatrix}\text{MeV}.
\end{align}
\end{subequations}
}}
Other possibility that works well is the following numerical five-texture zeros in the two-zero family case.
{
{
\begin{subequations}
\label{4.22a}
\begin{align}
 M_u^{\prime}&=\begin{pmatrix}
      0&0&123.038 - 285.496 i\\
0&1430.03& 18632.8 - 2336.25 i\\
123.038 + 285.496 i&18632.8+ 2336.25 i& 170033
     \end{pmatrix}\text{MeV},\\
M_d^{\prime}&=\begin{pmatrix}
     0& 13.2473& 0\\
13.2473 & 0&425.817\\
0& 425.817& 2796.6
    \end{pmatrix}\text{MeV},
\end{align}
\end{subequations}
}}
with $\lambda_{2u}=-m_c$.
\subsection{One-zero family}
A typical representation of this family is given by
\begin{equation}
\label{78}
 M_{u,d}=\begin{pmatrix}
      0& |B_{u,d}|& 0\\
|B_{u,d}|& C_{u,d}& 0\\
0& 0& A_{u,d}
     \end{pmatrix}.	
\end{equation}
The mass matrix $M_{u,d}$ is diagonalized as follows
\begin{align}
 O_{u,d}^\dag M_{u,d}O_{u,d}&=O_{u,d}^\dag\begin{pmatrix}
      0& |B_{u,d}|& 0\\
|B_{u,d}|& C_{u,d}& 0\\
0& 0& A_{u,d}
     \end{pmatrix}O_{u,d}\:,\\
&=
\begin{pmatrix}
 \lambda_{1{u,d}}&&\\
&\lambda_{2{u,d}}&\\
&&\lambda_{3{u,d}}
\end{pmatrix},
\end{align}
%
The following matricial functions allows us to write the elements of $M_{u,d}$ in terms of its eigenvalues  $\lambda_{i{u,d}}$. They are
%
\begin{align}
 \textrm{tr}M_{u,d}&=A_{u,d}+C_{u,d}=\lambda_{1{u,d}}+\lambda_{2{u,d}}+\lambda_{3{u,d}}\:,\\
\begin{split}
\textrm{tr}M_{u,d}^2&=A_{u,d}^2+2|B_{u,d}|^2+C_{u,d}^2\:,\\
&=\lambda_{1{u,d}}^2+\lambda_{2{u,d}}^2+\lambda_{3{u,d}}^2\:,
\end{split}
\\
\textrm{det}M_{u,d}&=-A_{u,d}|B_{u,d}|^2=\lambda_{1{u,d}}\lambda_{2{u,d}}\lambda_{3{u,d}}\:,
\end{align}
%
from which we have various solutions
\begin{description}
 \item[a)]
\begin{equation}
\label{99a}
\begin{split}
  A_{u,d}&=\lambda_{1{u,d}},\quad|B_{u,d}|=\sqrt{-\lambda_{2{u,d}}\lambda_{3{u,d}}}\:,\\
 C_{u,d}&=\lambda_{2{u,d}}+\lambda_{3{u,d}}\:,
\end{split}
\end{equation}

 \item[b)]

\begin{equation}
\label{100a}
\begin{split}
  A_{u,d}&=\lambda_{2{u,d}},\quad |B_{u,d}|=\sqrt{-\lambda_{1{u,d}}\lambda_{3{u,d}}}\:,\\
\quad C_{u,d}&=\lambda_{1{u,d}}+\lambda_{3{u,d}}\:,
\end{split}
\end{equation}

 \item[c)]
\begin{equation}
\label{101a}
\begin{split}
  A_{u,d}&=\lambda_{3{u,d}},\quad |B_{u,d}|=\sqrt{-\lambda_{1{u,d}}\lambda_{2{u,d}}}\:,\\
 C_{u,d}&=\lambda_{1{u,d}}+\lambda_{2{u,d}}\:.
\end{split}
\end{equation}
 \end{description}
Each one of these former cases were analyzed. Both representations {\it u-diagonal} and {\it d-diagonal} were worked. The Eqs.~(\ref{99a}, \ref{100a}, \ref{101a}), gives two possibilities for each case a), b) and c), depending of what eigenvalue is negative. In turn, each one of these cases, contain three possibilities depending of the negative eigenvalue assigned for the down~(up) mass matrix . In total there are 36 possibilities.
 Neither of this  cases were able to give models with five-texture or six-texture zeros.


\section{Analytical Five-Texture Zeros and the CKM Matrix}
\label{sV}
The five-texture zeros form of Eq.~(\ref{4.22a}), derived under the conditions given in section~\ref{s4a2}, is specially interesting because with the latest low energy data shows that it is a viable model, something not considered or rule out in papers like~\cite{b0,b4,c_6}. Therefore, let us assume the following five-texture zeros model
\begin{equation}
\label{5.1}
 M_u=
P^\dag\begin{pmatrix}
 0&0&|C_u|\\
0&A_u&|B_u|\\
|C_u|&|B_u|&\tilde B_u
\end{pmatrix}P,
\quad
M_d=
\begin{pmatrix}
 0&|C_d|&0\\
|C_d|&0&|B_d|\\
0&|B_d|&A_d
\end{pmatrix},
\end{equation}
where up and down quark mass matrices are given in the most general way, 
$P=\textrm{diag}(e^{-i\phi_{c_u}},e^{-i\phi_{b_u}},1)$ with $\phi_{b_u}\equiv 
\arg(B_u)$ and  $\phi_{c_u}\equiv \arg(C_u)$, where the phases for $M_d$ were no  considered because they can be absorbed, through a WB transformation, into $P$. Considering {$\lambda_{2u}=-m_c$},  we have from~(\ref{3.18}) through (\ref{35a}) that
{
\begin{equation}
 \begin{split}
  \tilde{B}_u&=m_u+m_t-m_c-A_u,\\
|B_u|&=\frac{\sqrt{A_u+m_c}\,\sqrt{m_t-A_u}\,\sqrt{A_u-m_u}}{\sqrt{A_u}},\\
|C_u|&=\frac{\sqrt{m_c}\,\sqrt{m_t}\,\sqrt{m_u}}{\sqrt{A_u}},
 \end{split}
\end{equation}}
where~(\ref{3.27}) was considered.

Taking into account ~(\ref{4.9}) and~(\ref{4.10}), for the down mass matrix  we have  that
\begin{equation}
 \begin{split}
  A_d&=m_d+m_b-m_s,\\
|B_d|&=\frac{\sqrt{m_d+m_b}\,\sqrt{m_b-m_s}\,\sqrt{m_s-m_d}}{\sqrt{m_d+m_b-m_s}},\\
|C_d|&=\frac{\sqrt{m_b}\,\sqrt{m_d}\,\sqrt{m_s}}{\sqrt{m_d+m_b-m_s}}.
 \end{split}
\end{equation}
The unitary matrix $U_u$ which diagonalizes $M_u$ is given by
{
\begin{widetext}
{
\begin{equation}
\label{5.4}
\begin{split}
 U_u=P^\dag\cdot p_2\cdot O_u\approx 
\begin{pmatrix}
\frac{\sqrt{Au-m_u}{e}^{i\,(\phi_{c_u}+x_u)}}{\sqrt{Au}} & -\frac{\sqrt{A_u+m_c}\,\sqrt{m_u}{e}^{i\,(\phi_{c_u}+y_u)}}{\sqrt{A_u}\,\sqrt{m_c}} & \frac{\sqrt{m_c}\,\sqrt{m_t-A_u}\,\sqrt{m_u}{e}^{i\,(\phi_{c_u}+z_u)}}{\sqrt{A_u}\,m_t}
\cr
 -\frac{\sqrt{A_u+m_c}\,\sqrt{m_t-A_u}\,\sqrt{m_u}\,{e}^{i(\phi_{b_u}+x_u)}}{\sqrt{A_u}\,\sqrt{m_c}\,\sqrt{m_t}} & -\frac{\sqrt{m_t-A_u}\sqrt{A_u-m_u}\,{e}^{i(\phi_{b_u}+y_u)}}{\sqrt{m_t}\sqrt{A_u}} & \frac{\sqrt{A_u+m_c}\sqrt{A_u-m_u}\,{e}^{i(\phi_{b_u}+z_u) }}{\sqrt{m_t}\sqrt{A_u}}
\cr 
\frac{\sqrt{A_u-m_u}\,\sqrt{m_u}\,e^{ix_u}}{\sqrt{m_c}\,\sqrt{m_t}} & \frac{\sqrt{A_u+m_c}\,e^{iy_u}}{\sqrt{m_t}} & \frac{\sqrt{m_t-A_u}\,e^{iz_u}}{\sqrt{m_t}}
\end{pmatrix},
\end{split}
\end{equation}}
\end{widetext}}
\noindent
where an additional phase factor $e^{iz_u}$ in third column of $O_u$~(Eq.~(\ref{32})) were added, in order to reproduce all phases present in the CKM matrix. The $3\times3$ matrix $p_2=
[(1,0,0),
(0,0,1),
(0,1,0)]
$ and the hierarchy~(\ref{19d}) together with~(\ref{3.27}) were considered.

And the unitary matrix $U_d$ which diagonalizes $M_d$ is given by
\begin{equation}
\label{5.5}
 U_d\approx\begin{pmatrix} 
 e^{ix_d}& -\frac{\sqrt{m_d}\,e^{iy_d}}{\sqrt{m_s}} & \frac{\sqrt{m_d}\,m_s}{(m_b)^{3/2}}\cr 
\frac{\sqrt{m_d}\,e^{ix_d}}{\sqrt{m_s}} & e^{iy_d}& \frac{\sqrt{m_s}}{\sqrt{m_b}}\cr 
-\frac{\sqrt{m_d}\,e^{ix_d}}{\sqrt{m_b}} & -\frac{\sqrt{m_s}\,e^{iy_d}}{\sqrt{m_b}} & 1
\end{pmatrix},
\end{equation}
where, in the process, a phase factor in the third column was not necessary to be included.  Now, we can easily find the CKM matrix $V=U_u^\dag U_d$. In particular, using the matrix form~(\ref{5.4}) and~(\ref{5.5}) for $U_u$, $U_d$ respectively, can
 survive current experimental tests. To leading order, we obtain.
{
{\footnotesize
\begin{subequations}
\label{5.6a}
\begin{align}
\label{5.6}
 |V_{ud}|&\approx|V_{cs}|\approx|V_{tb}|\approx 1,
\\
\label{5.7}
|V_{us}|\approx|V_{cd}|&\approx\left| \sqrt{\frac{A_u+m_c}{A_u}}\sqrt{\frac{m_u}{m_c}}
+{e}^{\pm i(\phi_{b_u}-\phi_{c_u})}\sqrt{\frac{m_d}{m_s}}\right|,
\\
|V_{cb}|\approx|V_{ts}|&\approx \left|\sqrt{\frac{m_s}{m_b}}-e^{\pm i\phi_{b_u}}\sqrt{\frac{A_u+m_c}{m_t}}\,\right|,
\\
\label{5.9}
\frac{|V_{ub}|}{|V_{cb}|}&\approx \sqrt{\frac{m_u}{m_c}}\left|\frac{\sqrt{\frac{A_u}{m_t}}-{e}^{-i\phi_{b_u}}\sqrt{\frac{A_u+m_c}{A_u}}\,\sqrt{\frac{m_s}{m_b}}}{\sqrt{\frac{A_u+m_c}{m_t}}-{e}^{-i\phi_{b_u}}\sqrt{\frac{m_s}{m_b}}}\right|,
\\
\label{5.10}
\frac{|V_{td}|}{|V_{ts}|}&\approx\sqrt{\frac{m_d}{m_s}},
\end{align}
\end{subequations}}}
\hspace{-1.5mm}where we assume that {$ m_u\ll A_u\ll m_t$}. The sign ``$+$'' for $V_{us},V_{cb}$ and ``$-$'' for $V_{cd},V_{ts}$. Note that if $A_u\gg m_c$ then $\frac{|V_{ub}|}{|V_{cb}|}\approx \sqrt{\frac{m_u}{m_c}}$, but this is not our case.

It is obvious that Eq.~(\ref{5.6}),~(\ref{5.7}) and~(\ref{5.10}) are consistent with the previous results~\cite{b4,b7}. A good fit of Eqs.~(\ref{5.6a}) and the CKM to the experimental data suggests
{
\begin{equation}
\begin{split}
 A_u&=1430.03\,\textrm{MeV},\;\phi_{b_u}=-0.124733, \;\phi_{c_u}=-1.16389,
 \\
  x_u=-1.83392,\; y_u&=-2.68335,\; z_u=0.00200664,\; x_d=-3.00697,\; y_d=0.344676,
 \end{split}
\label{5.7a}
\end{equation}}
\noindent
which  differ from the values given in~\cite{b4,b7}, {$\phi_1\approx\pi/3\sim(\phi_{b_u}-\phi_{c_u})$, such that it is an important contribution term of CP-violation in the context of present mass matrices}, and {$\phi_2\approx\pi/25\sim-\phi_{b_u}\rightarrow0$}. 
The numerical analysis shows that, by plugging for the quark masses the values given in~(\ref{17a}) and  the input parameters in~(\ref{5.7a}), we obtain the following absolute values for the mixing matrix
\begin{equation}
\label{5.8a}
 |V_{ckm}|=\begin{pmatrix}
 0.993&  0.255\pm0.030&0.00334\pm0.00094&
\\
0.255\pm 0.030& 1.004& 0.034\pm0.014
\\
0.0079\pm0.0020& 0.035\pm0.014& 1.011
 \end{pmatrix},
\end{equation}
in good agreement with the experimental measured values presented in~(\ref{2a}). For the Wolfenstein parameters  we find that
{
\begin{equation}
\label{5.9b}
 \begin{split}
  \lambda'&=0.247\pm0.027,\quad A'=0.55^{+0.43}_{-0.31},\\
\bar\rho'&=0.117\pm0.061,\quad\bar\eta'=0.361\pm0.070,
 \end{split}
\end{equation}}
which is in quite good agreement with the fit experimental values~(\ref{3.4}).
The inner angles of the CKM unitarity triangle, $V_{ud}V_{ub}^*+V_{cd}V_{cb}^*+V_{td}V_{tb}^*=0$, are
{
\begin{equation}
\label{5.8}
 \begin{split}
  \beta&=\arg\left(-\frac{V_{cd}V_{cb}^*}{V_{td}V_{tb}^*}\right)=24.4114^\circ,
\\
\alpha&=\arg\left(-\frac{V_{td}V_{tb}^*}{V_{ud}V_{ub}^*}\right)=82.6294^\circ,
\\
\gamma&=\arg\left(-\frac{V_{ud}V_{ub}^*}{V_{cd}V_{cb}^*}\right)=72.9592^\circ,
 \end{split}
\end{equation}}
tha are into the constraint stablished by~\cite{b1}.
And the Jarlskog invariant obtained is
{
\begin{equation}
\label{5.9a}
J'=\textrm{Im}(V_{us}V_{ub}^*V_{cs}^*V_{cb})=2.8322\times10^{-5},
\end{equation}}
which can be found into the interval given in~(\ref{3.7}).


\section{Conclusions}
\label{sVI}
Within the Standard Model framework, we have investigated texture zeros for quark mass matrices that reproduce the quark masses and experimental mixing parameters. To simplify the problem, without loss of generality, we consider that the quark mass matrices are Hermitian, since the right chirality fields are singlets under the  gauge symmetry SU(2). So, {for any model where the fields are right chiral singlet under the local gauge symmetry, we may consider that their mass matrices are Hermitian}. Specific six-texture zeros in quark mass matrices, including the Fritzsch model~\cite{c_1} and others like~\cite{b_2}, have already been discarded because they can not adjust their results to the experimental data known at present. In Sect.~\ref{sII}, together with the definition of WB transformation, it is  shown that the number of non-equivalent representations for the quark mass matrices is finite, which greatly simplifies the problem. 
Through WB transformations was relatively easy to find non-parallel four-texture  zeros  mass matrices. More difficult, but feasible, was the case for parallel four-texture zeros  mass matrices, {which were found in an exact way}. Significant was the consistent five-texture zeros quark mass matrix  found by us. Similarly, we show the impossibility, under any circumstances, to find mass matrices with six-texture zeros consistent with experimental data. This is a generalization of six-texture zeros  mass matrices discarded  by Fritzsch et al.

Throughout this letter, into the SM, we have used the fact that  all WB are equivalent. The opposite case is valid too, i.e., two  quark mass matrices representations giving the same physical quantities must be related through a WB transformation. Which is condensed in statement~(\ref{2.11a}).

By making appropriated WB transformations, numerical parallel and non parallel  four-texture zeros were found. An exhaustive deduction process allows us to find a five-texture zeros numerical structure compatible with the experimental data, Eqs.~(\ref{4.22}) and~(\ref{4.22a}).
This representation was found in the two-zero family case. Equivalent representations are given in Table~\ref{ta1}.

We have determined the impossibility to find quark mass matrices having a total of six-texture zeros which are consistent with the measured values of the quark masses and mixing angles. While, a consistent model with five-texture zeros were successful. The five texture zero Ansatz of Eq.(\ref{5.1})~(with $\lambda_{2u}=-m_c$), together with some assumptions
which include appropriated values for $A_u, \phi_{b_u}$, $\phi_{c_u}$, $x_u$, $y_u$, $z_u$, $x_d$ and $y_d$ does lead to successful predictions for $V_{CKM}$ such
as those of Eqs.(\ref{5.6a}), (\ref{5.8a}), (\ref{5.9b}), (\ref{5.8}) and~(\ref{5.9a})~\footnote{{In the case $\lambda_{1u}=-m_u$ similar results can be found.}}.  One nice thing about five-texture zeros quark mass matrices~(\ref{5.1}) is that no hierarchies on quark masses  is necessary to be imposed to make correct predictions, although, expressions~(\ref{5.6a}) become in a more complex notation. 

\section*{Acknowledgments}

This work was partially supported by VIPRI in the U. de Nari\~no, approval Agreement Number 20.

\appendix
\section{Verification of the Method.}
\label{a2}
%
The paper~\cite{b0} uses the following quark mass data:
{\small
\begin{align}
 m_u&=2.50~\textrm{MeV}, m_c=600~\textrm{MeV}, m_t=174000~\textrm{MeV}, \\
m_d&=4.00~\textrm{MeV},m_s=80~\textrm{MeV}, m_b=3000~\textrm{MeV}.
\end{align}}
and the numerical CKM matrix used is
\begin{widetext}
{
\begin{equation}
V=
\begin{pmatrix}
 0.036195+0.97493 i & -0.057798+0.21177 i& 0.00037188 - 0.0035669 i \\
-0.21247 + 0.054471 i & 0.97351 + 0.050582i & -0.0044010 - 0.039760i \\
0.0043605 + 0.0083871 i & 
0.0086356 - 0.038067 i &
0.99836 + 0.040693 i
\end{pmatrix}.
\end{equation}
}
\end{widetext}
We assume the following case:
%
\begin{align}
\lambda_{1u}&=-m_u,\lambda_{2u}=m_c,\lambda_{3u}=m_t,\\
 \lambda_{1d}&=-m_d,\lambda_{2d}=m_s,\lambda_{3d}=m_b.
\end{align}
Then, the quark mass matrices~(\ref{3}) are
{\scriptsize
\begin{align}
\label{49}
 M_u&=\begin{pmatrix}
  -2.5&&\\
&600&\\
&&174000
 \end{pmatrix}\textrm{MeV},\\
\label{50_4}
M_d&=\begin{pmatrix}
 0.086447 &  -3.4055 + 17.655i & -0.039835 - 10.774i\\
-3.4055 - 17.655i & 80.631 & -17.515 - 115.56i\\
-0.039835 + 10.774i & -17.515 + 115.56i & 2995.3
\end{pmatrix}\textrm{MeV}.
\end{align}}
Let us use the diagonalization matrix~(\ref{32}) with  $x=\pi$ and $y=\pi$.
\begin{widetext}
\begin{equation}
{
 O_u\approx 10^{-3}
\begin{pmatrix}
 -997.92 & -64.527 \sqrt{\frac{A_u-600}{A_u}} & 0.22297
   \sqrt{\frac{174000-A_u}{A_u}} \\
&&\\[-3mm]
 0.15442 \sqrt{A_u} & -2.3965 \sqrt{A_u-600} &2.4014 \sqrt{174000-A_u}
   \\
&&\\[-3mm]
 -0.15442 \sqrt{\frac{(174000-A_u) (A_u-600)}{A_u}} & 2.3965 \sqrt{174000-A_u} &
   2.4014 \sqrt{A_u-600}
\end{pmatrix}}
\end{equation}
\end{widetext}
where the approximation $A_u>>m_u$ was assumed because of the restriction~(\ref{40}). The matrix $O_u$ now  plays the role of a unitary matrix to make the WB transformation on~(\ref{49}) and~(\ref{50_4}). The entries, in the new representation, depend of $A_u$. In order to have a  texture zeros at the entry $(1,3)$, we need to solve
\begin{widetext}
{\footnotesize
\begin{equation}
\begin{split}
 M_d(1,3)=Y(A_u)&\propto
(8144.2- 42221 i) A_u\sqrt{174000-A_u}+(95.463+ 25819 i) A_u
\sqrt{A_u-600}
\:-\\
&(10852)\sqrt{A_u(A_u-600)(174000-A_u)} -(9.3588 - 61.745 i)
   (174000-A_u)\sqrt{A_u}\:+
\\&(2714.0+ 17906  i)(A_u-600) \sqrt{A_u}
 +(0.0013716 - 0.37097 i) (174000-A_u)\sqrt{A_u-600}\\
&-(33.934 + 175.92i)(A_u-600)\sqrt{174000-A_u}\approx0,
\end{split}
\end{equation}}
\end{widetext}
%
whose solution is 
$
 A_u\approx 84621~\textrm{MeV},
$
%
which agrees perfectly with the value given in the aforementioned paper. The quark mass matrices~(\ref{49}) and~(\ref{50_4}) take the form
%
{\footnotesize
\begin{align}
 M_u^{\prime}&=O_uM_uO_u^T=
\begin{pmatrix}
 0 & 55.537 & 0\\
55.537 & 89977 & 86660\\
0 & 86660 & 84621
\end{pmatrix}\textrm{MeV},\\
M_d^{\prime}&=O_uM_dO_u^T\\
&\approx
\begin{pmatrix}
 0 & 2.5792+25.325i & 0\\
2.5792-25.325i & 1600.5 & 1456.0+114.63i\\
0 & 1456.0-114.63i & 1475.5
\end{pmatrix}\text{MeV}.	
\end{align}}
At the present stage we have not yet obtained the matrices given in (25) and (26) of paper~\cite{b0}. But we can make an additional WB transformation using the following  diagonal unitary matrix with phase entries
\begin{equation}
 P=\begin{pmatrix}
    1&&\\
&e^{i 4.4984}&\\&&e^{-i 0.063300} 
   \end{pmatrix}.
\end{equation}
We finally get the desired matrices
{\footnotesize
\begin{align}
\begin{split}
 M_u^{\prime\prime}&=P^\dag M_u^{\prime}P\\
&=
\begin{pmatrix}
 0 & -11.794-54.270i& 0\\ 
-11.794+54.270i & 89977 & -13009+85678i\\
0&-13009-85678i& 84621
\end{pmatrix}\textrm{MeV}, 
\end{split} \\
\begin{split}
M_d^{\prime\prime}&=P^\dag M_d^{\prime}P\\
&=
\begin{pmatrix}
 0& 24.199-7.8983i &0\\
 24.199+7.8983i& 1600.5 & -331.91+1422.3i\\
0& -331.91-1422.3i& 1475.5
\end{pmatrix}\textrm{MeV}.
\end{split}
\end{align}}



\end{document}